\documentclass{appolb}
\usepackage{epsfig}
\begin{document}
\title{Extension of effective lagrangian approach to structure of selected nuclei far from stability.}
\author{Marco A. Huertas
\address{Department of Physics, College of William and Mary}
}
%\and
%the Name(s) of other Author(s)
%\address{and their affiliation}
\maketitle
\begin{abstract}
%Here comes the abstract
In a previous paper, the convergence of a new effective field theory and density functional theory (EFT/DFT) approach to the description of the nuclear many-body system was studied. The most sophisticated parameter set (here G1) determined by Furnstahl, Serot and Tang from a fit along the valley of stability was found to provide quantitative predictions for total binding energies and single-particle and single-hole binding energies, spins, and parities for selected doubly-magic nuclei far from stability. Binding energies of all the even Sn isotopes (Z=50) were also well described. Here the calculations are extended to all isotones with magic numbers N=28, 50, 82, 126, and isotopes with Z=28, 50, 82 with similar predictive results.
\end{abstract}

\section{Introduction.}
In a previous paper~\cite{marco1} the convergence of a new approach to the description of the nuclear many-body system, based on principles of effective field theory (EFT) and density functional theory (DFT)~\cite{f.s.t.}, was studied by applying it to selected nuclei far from stability. The results of the calculations showed an overall agreement with experiment below 1\% for the total binding energy of the doubly magic nuclei $^{132}_{\ 50}\textrm{Sn}_{82}$, $^{100}_{\ 50}\textrm{Sn}_{50}$, $^{48}_{28}\textrm{Ni}_{20}$ and $^{78}_{28}\textrm{Ni}_{50}$ as well as for the entire range of known even-even Sn isotopes. In addition, calculations for the chemical potential of nuclei differing by one particle or one hole from these doubly-magic ones, produced results with an overall agreement with experimental values below the 10\% level, and also correctly predicted the spin and parity of these nuclei. It was also shown in~\cite{marco1} that the calculations converged to the experimental values as the level of approximation of the effective lagrangian was increased. From the various levels of approximation used, it was found that the ones with the highest levels reproduced better the properties of nuclei far from stability. In addition to this, it was also observed that when this approach was tested to the whole range of even-even Sn isotopes the results obtained agreed equally well with the experimental values.

Based on the agreement of these results, a calculation of $\beta$-transition rates was performed for nuclei close to $^{132}_{\ 50}\textrm{Sn}_{82}$~\cite{marco2}. Here the objective was to test, through a semi-leptonic process, the accuracy of the single-particle Dirac wave functions obtained in~\cite{marco1} describing a particle or a hole outside the filled core. The calculations where done using a current-current interaction hamiltonian in which the lepton current is treated in first-order perturbation theory while the nuclear currents are expanded in multipoles and treated to all orders. The nuclear electroweak currents used in these calculations are those derived directly from the same effective lagrangian describing the nuclear many-body system. In this way the calculations are done based on a self-consistent approach. Two different cases where explored in~\cite{marco2}: particle-particle transitions $^{133}_{\ 50}$Sn$_{83}$ $\rightarrow$ $^{133}_{\ 51}$Sb$_{82}+e^{-}+\overline{\nu}_{e}$ and hole-hole transitions $^{131}_{\ 49}$In$_{82}$$\rightarrow$$^{131}_{\ 50}$Sn$_{81}+e^{-}+\overline{\nu}_{e}$. The results obtained indicated that the particle-particle transitions proceeding between ground states were well reproduced (within the 5\% level) while transitions to or from excited states deviate systematically from the experimental values. This behavior was expected based on the fact that ground-state observables (in this case the ground-state density and therefore presumably the wave functions) are best reproduced on the basis of DFT. Results obtained in the hole-hole case, in which all transitions start or end on an excited state, are less reliable in reproducing the $\beta$-transition rates. In any case, more studies are underway to understand this behavior.

Both papers showed that this EFT/DFT approach can be used successfully to describe \textit{some} ground-state properties of the selected nuclei far from stability; this extends the applicability of the approach beyond the valley of stability. To further explore the range of validity of this approach, the calculations done in~\cite{marco1} are extended here to all isotones with magic numbers N= 28, 50, 82, 126 and isotopes with Z= 28, 82, 126. Extensions of the calculations made in~\cite{marco2} are considered in another paper.

The calculations done in~\cite{marco1}, which are extended here, are based on an energy functional of the ground-state density $E[\rho_{g.s.}]$. By minimization of this energy functional with respect to the density one obtains the ground-state energy and density of the nuclear system. The existence of a unique functional of the density is guaranteed by the Hohenberg-Kohn theorem~\cite{dft1,dft2}. This minimization process can be computed using a procedure developed by Kohn and Sham in which the ground-state density of a interacting fermion system is computed using the single-particle wave functions obtained from an equivalent systems of non-interacting fermions subject to an \textit{appropriate} external potential, known as the Kohn-Sham potential~\cite{dft1,dft2}. The basis for this procedure is contained in the Hohenberg-Kohn theorem which can be stated by saying that there is a one-to-one correspondence between the ground-state density of a system of interacting fermions subject to an external potential and the external potential~\footnote{In relativistic theories the energy functional is written in terms of more than one density: e.g. scalar densities $\bar{\psi}\psi$, vector densities $\bar{\psi}\gamma_{\mu}\psi$, etc.}.

Although the formulation based on DFT is straightforward, the exact energy functional is unknown, particularly in the nuclear system where the exact form of the nuclear interaction is still a subject of intense research. To overcome this, an energy functional is constructed based on the principles of effective field theory (EFT)~\cite{f.s.t.}. The starting point is an effective lagrangian density that satisfies the symmetries imposed by QCD, which is considered the underlying theory of the nuclear dynamics. This lagrangian is built using the lowest lying hadronic degrees of freedom ( as well as the electromagnetic field). The result is the most general lagrangian containing (in principle an infinite number of) interaction terms constructed using these fields and subject to the imposed symmetries. The various interaction terms in this lagrangian are ordered in powers of the fields and their derivatives. This expansion is equivalent to an expansion in the ratios of the fields to the nucleon mass and the Fermi momentum to the nucleon mass. These expansion parameters are small: $\approx$1/3 and $\approx$1/4 respectively. Naive Dimensional Analysis (NDA), which is a prescription by which the various interaction terms are suppressed by powers of a relevant mass scale $\Lambda$ (the nucleon mass M in this case), indicates that the remaining coupling constants should be of order unity, a characteristic called ''naturalness.'' This natural size of the constants can only be verified after fitting the theory to experimental data. This procedure was done in~\cite{f.s.t.} where it is shown that the remaining constants are indeed ''natural.'' There the constants were fit to ground-state properties of selected spherical nuclei along the valley of stability. The effective lagrangian, as well as the corresponding equations of motion obtained from it, are shown in the next section.

As has been mentioned before, this paper explores the extent of applicability of this EFT/DFT approach. To understand the objectives of this work, one has to keep in mind that the coupling constants of the effective lagrangian have been fit to properties of selected \textit{stable} nuclei lying in the $\beta$-stability region. Therefore the results of~\cite{marco1}, and the ones presented here, test how well this theory reproduces ground-state properties \textit{outside} the stability region. As indicated in~\cite{marco1} the relevant observables which can be reproduced, based on DFT, are the \textit{ground-state binding energy} and \textit{chemical potential} of the nuclear system, so this paper concentrates on these quantities. The calculations focus on nuclei with semi-magic number of neutrons, i.e. N = 28, 50, 82 and 126 or protons, i.e. Z = 28, 50 and 82. In general it is seen that the predictions of the total ground-state binding energy, for both isotones and isotopes, lie below the 1\% level as was found in~\cite{marco1}, and that as the size of the nucleus increases, the error is reduced. The agreement between the results of the chemical potential with experimental data is also consistent with the conclusions reached in~\cite{marco1}. In the case of the prediction of the ground-state J$^{\pi}$ values of odd nuclei neighboring even-even  ones, the agreement with experimental values varies between isotones and isotopes. When the number of neutrons gets very large in the regime N$>$Z, the level density gets very high and the detailed ordering of the levels between mayor shells depends sensitively on the details of the various densities. This small energy shifts ($\approx$ tens keV) make the predictions of the ground-state quantum numbers less accurate, though the chemical potentials are still relatively well-described.

All nuclei selected for this study are essentially spherical~\cite{deformations}, therefore the approximation of spherically symmetric nuclei, made here, is assumed to be an accurate one. The calculations are done using the most convergent parameter set obtained in~\cite{f.s.t.}, called G1, which corresponds to the most sophisticated effective lagrangian.

To summarize, this paper extends the calculations made in~\cite{marco1} to known semi-magic isotones and isotopes using the G1 parameter set. In the following section the formulation of the Kohn-Sham equations and potentials is described, and this is followed by a section on results. The paper ends with the conclusions and proposes new directions of investigation to further test this approach.

\section{Formalism.}
In this section an overview of the formalism of EFT/DFT is presented. A more thorough presentation can be found in~\cite{f.s.t.}.

The energy functional that describes the dynamics of the nuclear system is obtained from an effective lagrangian. This is constructed using the relevant lowest lying hadronic degrees of freedom~\cite{f.s.t.}, which are the pion and the rho fields (isovectors) and the nucleon field (isospinor). In addition to these, a scalar field is included to reproduced the known short-range attraction of the nuclear force, and a vector field (associated to the $\omega$ meson) that reproduces the mid-range repulsion of the nuclear interaction is also included.

The pion field is introduced as the phase of a chiral rotation of the identity matrix in isospin space. A field $\xi(x)$ is defined in the following way:
\[
\xi(x) = \textrm{exp}(i\pi(x)/f_{\pi}).
\]
where the pion field $\pi(x)$ is defined here by $\pi(x)=\frac{1}{2}\mathbf{\vec \tau \cdot \vec \pi}$.

Nucleons are included as a two-component isospinor field $N(x)$. The upper and lower components correspond to the proton and neutron fields, respectively.
\begin{equation}
N = \left( \begin{array}{c} p(x) \\ 
n(x) \end{array} \right)
\end{equation}

The $\rho_{\mu}$-field has been added to the lagrangian to include contributions coming from the asymmetry between the number of protons and neutrons.  In addition to the above fields, an electromagnetic field $A_{\mu}$ is included and the electromagnetic structure of the nucleons is also taken into account. Their anomalous magnetic moments, labeled as $\lambda_{p}$ and $\lambda_{n}$ for proton and neutron respectively, then enter. 

The effective lagrangian contains, in principle, an infinite number of terms. In order to make the theory predictive, these terms have to be arranged according to some ordering scheme that takes into account all equally contributing terms. Here the various interaction terms in the effective lagrangian are arranged following the prescriptions of NDA~\cite{NDA}. As mentioned before, all calculations performed here will be done using the parameter set G1 which corresponds to the most sophisticated lagrangian density of~\cite{f.s.t.}, and is reproduced in Table~\ref{g1-set}. The pion fields are omitted from the following lagrangian since they do not develop a mean field in the class of nuclei considered here.

\begin{table}%[H] add [H] placement to break table across pages
\caption{\label{g1-set} Numerical values of constants for the G1 parameter set~\cite{f.s.t.}.}
%\begin{ruledtabular}
\begin{center}
\begin{tabular}{|cc|cc|}

\hline
% Lines of table here ending with \\
Constant & G1 & Constant & G1\\
\hline
m$_{\textrm{s}}$ & 506.7 &  $\zeta_{0}$ & 3.5249 \\
g$^{2}_{\textrm{s}}$ & 97.39 & $\eta_{\rho}$ & -0.2722 \\
g$^{2}_{\textrm{v}}$ & 147.09 & $\alpha_{1}$ & 1.8549\\
g$^{2}_{\rho}$ & 77.033 &  $\alpha_{2}$ & 1.788 \\
$\eta_{1}$ & 0.0706 & f$_{\textrm{v}}$ & 0.4316 \\
$\eta_{2}$ & -0.96161 &  f$_{\rho}$ & 4.1572 \\
$\kappa_{3}$ & 2.2067 &  $\beta_{\textrm{s}}$ & 0.02844 \\
$\kappa_{4}$ & -10.090 & $\beta_{\textrm{v}}$ & -0.24992 \\
\hline
\end{tabular}
\end{center}
%\end{ruledtabular}
\end{table}

The nucleonic part of the lagrangian density is~\footnote{The metric and conventions used are those of~\cite{f.s.t.,serot-qhd}, and differ from~\cite{waleckabook}.}:
\begin{eqnarray}
\label{eq:Lnucleon}
\mathcal{L}_{N}(x) & = & \bar{N}(i\gamma^{\mu}D_{\mu}-M+g_{s}\phi)N -\frac{f_{\rho}g_{\rho}}{4M}\bar{N}\rho_{\mu\nu}\sigma^{\mu\nu}N -\frac{f_{V}g_{V}}{4M}\bar{N}\rho_{\mu\nu}\sigma^{\mu\nu}N \nonumber \\
 & & -\frac{e}{4M}F{\mu\nu}\bar{N}\lambda\sigma^{\mu\nu}N - \frac{e}{2M^{2}}\bar{N}\gamma_{\mu}(\beta_{s}+\beta_{V}\tau_{3})N\partial_{\nu}F^{\mu\nu}
\end{eqnarray}
where $D_{\mu}=\partial_{\mu}+ig_{\rho}\rho_{\mu}+ig_{V}V_{\mu}+\frac{i}{2}eA_{\mu}(1+\tau_{3})$.

The mesonic part is the following:
\begin{eqnarray}
\label{eq:Lmeson}
\mathcal{L}_{M}(x) & = & \frac{1}{2}(1+\alpha_{1}\frac{g_{s}\phi}{M})\partial_{\mu}\phi\partial^{\mu}\phi - \frac{1}{2}tr(\rho_{\mu\nu}\rho^{\mu\nu}) - \frac{1}{4}(1+\alpha_{2}\frac{g_{s}\phi}{M})V_{\mu\nu}V^{\mu\nu} \nonumber \\
 & & -\frac{e}{2g_{\gamma}}F_{\mu\nu}[tr(\tau_{3}\rho^{\mu\nu})+\frac{1}{3}V^{\mu\nu}]+\frac{1}{2}(1+\eta_{1}\frac{g_{s}\phi}{M}+\frac{\eta_{2}g^{2}_{s}\phi^{2}}{2 M^{2}})m^{2}_{V}V_{\mu}V^{\mu} \nonumber \\
 & & + \frac{1}{4!}\zeta_{0}g^{2}_{V}(V_{\mu}V^{\mu})^{2}+(1+\eta_{\rho}\frac{g_{s}\phi}{M})m^{2}_{\rho}tr(\rho_{\mu}\rho^{\mu}) \nonumber \\
 & & - m^{2}_{s}\phi^{2}(\frac{1}{2}+\frac{\kappa_{3}g_{s}\phi}{3!M}+\frac{\kappa_{4}g^{2}_{s}\phi^{2}}{4!M^{2}}) - \frac{1}{4} F^{\mu \nu}F_{\mu \nu}
\end{eqnarray}
 
In the above lagrangian, $\phi$ is the scalar field, $V_{\mu}$ the vector field, $\rho_{\mu}$ the isovector field rho defined by $\rho_{\mu}=\frac{1}{2}\vec{\tau}\cdot \vec{\rho}_{\mu}$ and $A_{\mu}$ is the electromagnetic field. 

From the above effective lagrangian density one obtains the Kohn-Sham equations and the auxiliary equations for the potentials. To do so, the various mesonic and electromagnetic fields as considered as local classical fields. Then the necessary equations are obtained by calculating the Euler-Lagrange equations. Since the nuclei considered here are (essentially) spherical~\cite{deformations}, the various vector fields: $V_{\mu}$, $\rho_{\mu}$ and $A_{\mu}$, develop only their time-components $V_{0}$, $\rho_{0}$ and $A_{0}$ respectively. In addition, the isovector $\rho_{0} (x)$-field develops only its zero-charged component denoted here as $b_{0}$.

The Dirac hamiltonian for the nucleon fields as derived from the effective lagrangian takes the form:
{\setlength\arraycolsep{2pt}
\begin{eqnarray}
\label{eq:Dirac-equation}
h(x) & = &  -i\mathbf{\vec{\alpha} \cdot \vec{\nabla}} + W(x) + \frac{1}{2}\tau_{3}R(x) + \beta(M-\Phi(x)) {}+\frac{1}{2}(1+\tau_{3})A(x){} \nonumber \\
& & {}-\frac{i}{2M} \beta \vec{\alpha} \cdot (f_{\rho} \frac{1}{2} \tau_{3} \vec{\nabla} R + f_{v} \vec{\nabla} W) {}+\frac{1}{2M^{2}} (\beta_{s} + \beta_{v} \tau_{3})\nabla^{2} A{}\nonumber \\
& & {} - \frac{i}{2M} \lambda \beta \mathbf{\vec{\alpha} \cdot \vec{\nabla}} A       {}
\end{eqnarray}}
Here $\lambda=\frac{1}{2}\lambda_{p}(1+\tau_{3})+\frac{1}{2}\lambda_{n}(1-\tau_{3})$ and the numerical values used for the anomalous magnetic moments are $\lambda_{p}=1.793$, $\lambda_{n}=-1.913$.

The mean meson and electromagnetic fields are denoted by $W=g_{v}V_{0}$, $\Phi=g_{s}\phi_{0}$, $R=g_{\rho}b_{0}$, and $A=eA_{0}$ respectively. The quantities ${f_{\rho},f_{v},\beta_{s},\beta_{v}}$ are parameters fit to experiment.

The equations describing the various meson and electromagnetic fields are the following:
{\setlength\arraycolsep{2pt}
\begin{eqnarray}
\label{eq:Phi-equation}
- \nabla^{2}\Phi + m_{s}^{2}\Phi & = & g_{s}^{2}\rho_{s}(x) - \frac{m_{s}^{2}}{M}\Phi^{2} (\frac{\kappa_{3}}{2} + \frac{\kappa_{4}}{3!}\frac{\Phi}{M}){}+\frac{g_{s}^{2}}{2M}(\eta_{1} + \eta_{2} \frac{\Phi}{M}) \frac{m_{v}^{2}}{g_{v}^{2}}W^{2} {} \nonumber \\
& & +\frac{g_{s}^{2} \eta_{\rho}}{2M} \frac{m_{\rho}^{2}}{g_{\rho}^{2}} R^{2} + \frac{\alpha_{1}}{2M} [(\vec{\nabla} \Phi)^{2} + 2 \Phi \nabla^{2} \Phi] {} + \frac{\alpha_{2} g_{s}^{2}}{2Mg_{v}^{2}} (\vec{\nabla} W)^{2} \nonumber \\
\end{eqnarray}}
Here $\rho_{s}$ is the baryon Lorenz scalar density and $g_{s}$, $m_{s}$, $\kappa_{3}$, $\kappa_{4}$, $\eta_{1}$, $\eta_{2}$, $g_{v}$, $g_{\rho}$, $\alpha_{1}$, $\alpha_{2}$ are again parameters fit to experiment. Furthermore,
{\setlength\arraycolsep{2pt}
\begin{eqnarray}
\label{eq:W-equation}
- \nabla^{2}W + m_{v}^{2}W & = & g^{2}_{v}[\rho_{B}(x)+\frac{f_{V}}{2M}\mathbf{\vec{\nabla}\cdot} (\rho^{T}_{B}(x)\mathbf{\hat{r}})] -(\eta_{1}+\frac{\eta_{2}}{2}\frac{\Phi}{M}) \frac{\Phi}{M}m^{2}_{v} W {}\nonumber \\
 & &{}-\frac{1}{3!}\zeta_{0}W^{3} + \frac{\alpha_{2}}{M}(\vec{\nabla} \Phi \cdot \vec{\nabla} W + \Phi \nabla^{2} W) - \frac{e^{2}g_{v}}{3g_{\gamma}}\rho_{\mathrm{chg}}(x) \nonumber \\
\end{eqnarray}}
Here $\rho_{B}$ is the baryon density, $\rho^{T}_{B}$ is the baryon tensor density, $\rho_{\mathrm{chg}}$ is the charge density and $f_{v}$, $\zeta_{0}$ are parameters. In addition,
{\setlength\arraycolsep{2pt}
\begin{eqnarray}
\label{eq:R-equation}
- \nabla^{2} R + m^{2}_{\rho} R & = &\frac{1}{2} g^{2}_{\rho}[\rho_{3}(x)+\frac{f_{\rho}}{2M}\mathbf{\vec{\nabla}\cdot} (\rho^{T}_{3}(x)\mathbf{\hat{r}})]-\eta_{\rho}\frac{\Phi}{M}m^{2}_{\rho}R-\frac{e^{2}g_{\rho}}{g_{\gamma}}\rho_{\mathrm{chg}}(x) \nonumber \\
\end{eqnarray}}
Here $\rho_{3}$ and $\rho^{T}_{3}$ are the isovector densities, $f_{\rho}$, $\eta_{\rho}$, $g_{\rho}$ are parameters and $g_{\gamma}$=5.01, is the coupling of the photon to the $\omega$-meson. Finally,
\begin{equation}
\label{eq:A-equation}
- \nabla^{2}A = e^{2}\rho_{\mathrm{chg}}(x) 
\end{equation}
where $\rho_{\mathrm{chg}}$ is the charge density. All parameters used in these equations are tabulated in Table~\ref{g1-set}.

From the above Dirac hamiltonian one can write down the energy eigenvalue equation which will give the single-particle wave functions that are used to construct the ground-state densities. The various mesons and electromagnetic fields play the role of the Kohn-Sham potentials. The whole set of equations (\ref{eq:Phi-equation}) - (\ref{eq:A-equation}) and (\ref{eq:Single-levels}), below, form the \textit{Kohn-Sham equations}.
\begin{equation}
\label{eq:Single-levels}
h\psi_{\alpha}(x) = E_{\alpha} \psi_{\alpha}(x)
\end{equation}

To obtain the single-particle wave functions from (\ref{eq:Single-levels}), they are expressed in terms of Dirac spherical wave functions: 
\begin{equation}
\label{eq:Nucleon-field}
\psi_{\alpha}(x) = 
\left( 
\begin{array}{c}
\frac{i}{r} G_{a}(r) \Phi_{\kappa m} \\
-\frac{1}{r} F_{a}(r) \Phi_{-\kappa m}
\end{array}
\right) \zeta_{t}
\end{equation}
where the $\Phi_{\kappa m}$ is a spin spherical harmonic. $\zeta_{t}$ is a two-component spinor and the index t is equal to 1/2 for protons and -1/2 for neutrons. By inserting this form of the solution into Eq.(\ref{eq:Single-levels}) and using Eq.(\ref{eq:Dirac-equation}) a set of two coupled first-order differential equations for the G and F functions is obtained:

\begin{equation}
\label{eq:G-equation}
\left(\frac{d}{dr} + \frac{\kappa}{r}\right)G_{a}(r) - [E_{a} - U_{1}(r) + U_{2}(r)]F_{a}(r) - U_{3}G_{a}(r) = 0
\end{equation}
\begin{equation}
\label{eq:F-equation}
\left(\frac{d}{dr} - \frac{\kappa}{r}\right)F_{a}(r) + [E_{a} - U_{1}(r) - U_{2}(r)]G_{a}(r) + U_{3}F_{a}(r) = 0
\end{equation}
where the single-particle potentials are given by
\begin{eqnarray}
U_{1}(r) & \equiv & W(r)+t_{a}R(r)+(t_{a}+\frac{1}{2})A(r){}\nonumber \\
 & & +\frac{1}{2M^{2}}(\beta_{s} + 2t_{a}\beta_{V})\nabla^{2}A(r) {} \\
U_{2}(r) & \equiv & M- \Phi(r) {}\\
U_{3}(r) & \equiv & \frac{1}{2M} \{ f_{V}W'(r)+t_{a}f_{\rho}R'(r){} \nonumber \\
& &{}+A'(r)[(\lambda_{p}+\lambda_{n})/2+t_{a}(\lambda_{p}-\lambda_{n})] \} {} 
\end{eqnarray}
Here the prime indicates a radial derivative; \textit{e.g.} $W^{`}(r)=dW(r)/dr$.
Once the single-particle wave functions are calculated, the various densities that appear on the r.h.s. of the meson equations can be obtained. They are defined as follows:
\begin{eqnarray}
\label{eq: scalar-dens}
\rho_{s}(x) & = & \sum_{\alpha}^{occ}\frac{2j_{a}+1}{4\pi r^{2}} \left( G^{2}_{a}(r)-F^{2}_{a}(r) \right) {} \\
\label{eq: vector-dens}
\rho_{B}(x) & = & \sum_{\alpha}^{occ}\frac{2j_{a}+1}{4\pi r^{2}} \left( G^{2}_{a}(r)+F^{2}_{a}(r) \right) {} \\
\rho^{T}_{B}(x) & = & \sum_{\alpha}^{occ}\frac{2j_{a}+1}{4\pi r^{2}}2G_{a}(r)F_{a}(r) {} \\
\rho_{3}(x) & = & \sum_{\alpha}^{occ}\frac{2j_{a}+1}{4\pi r^{2}} (2t_{a}) \left( G^{2}_{a}(r)+F^{2}_{a}(r) \right) {} \\
\rho^{T}_{3}(x) & = & \sum_{\alpha}^{occ}\frac{2j_{a}+1}{4\pi r^{2}} (2t_{a})2G_{a}(r)F_{a}(r){} 
\end{eqnarray}
where the sum goes over the occupied orbitals.
In addition to these, the charge density is defined as the sum of two pieces: a direct nucleon charge density $\rho_{d}(\mathbf{x})$  and the vector meson contribution $\rho_{m}(\mathbf{x})$
\begin{equation}
\label{eq:charge dens}
\rho_{\mathrm{chg}}\equiv \rho_{d}(\mathbf{x})+\rho_{m}(\mathbf{x})
\end{equation}

The direct part is given by
\begin{eqnarray}
\rho_{d}(x) &=& \rho_{p}(x)+ \frac{1}{2M} \mathbf{\vec{\nabla} \cdot} (\rho^{T}_{\mathrm{a}}(x)\mathbf{\hat{r}}) +\frac{1}{2M^{2}}[\beta_{s}\nabla^{2}\rho_{B}(x)+\beta_{v}\nabla^{2}\rho_{3}(x)]{}
\end{eqnarray}
and the vector meson contribution arising from the coupling of the neutral vector mesons to the photon takes the form
\begin{equation}
\rho_{m}(x)=\frac{1}{g_{\gamma} g_{\rho}} \nabla^{2}R+\frac{1}{3g_{\gamma} g_{v}} \nabla^{2}W
\end{equation}

Here the point proton density $\rho_{p}$ and nucleon tensor density $\rho^{T}_{\mathrm{a}}$ are given by
\begin{equation}
\rho_{p}=\frac{1}{2}(\rho_{B}+\rho_{3})
\end{equation}
\begin{equation}
\rho^{T}_{\mathrm{a}}=\sum^{occ}_{\sigma} \psi^{\dagger}_{\sigma}(x)i\lambda \beta \mathbf{\vec{\alpha} \cdot \hat{r}} \psi_{\sigma}(x)
\end{equation}

All the above equations were derived from the effective lagrangian given in Eqs. (\ref{eq:Lnucleon}) and (\ref{eq:Lmeson}), and are the same as those used in~\cite{marco1} and reproduce those shown in~\cite{f.s.t.}. To solve them an independently developed program was written to solve these local, coupled, non-linear equations (\ref{eq:Phi-equation})-(\ref{eq:A-equation}) and Eqs. (\ref{eq:G-equation}) and (\ref{eq:F-equation}), where the appropriate densities are defined in Eqs. (16) - (20). This program was used to obtain the results in~\cite{marco1} and reproduces the results shown in~\cite{f.s.t.}.

The meson equations are solved iteratively using a Greens function method. Because NDA guarantees that each additional term on the rhs of Eqs. (\ref{eq:Phi-equation}) - (\ref{eq:A-equation}) is smaller than the previous one, convergence is both expected and obtained.

The whole system of Eqs. (\ref{eq:G-equation}), (\ref{eq:F-equation}) and (\ref{eq:Phi-equation}) - (\ref{eq:A-equation}) is solved self-consistently until a global convergence is reached.

\section{Results.}
The procedure described above to solve the Kohn-Sham equations to find the eigenvalues and wave functions is performed using the parameter set G1 given in~\cite{marco1,f.s.t.}, and which is reproduced in Table~\ref{g1-set} for completeness. These equations are solved and used to calculate the ground-state total binding energy of the selected nuclei~\footnote{The total binding energy of the system follows directly from the effective lagrangian and hamiltonian, see~\cite{f.s.t.}}. Before starting the discussion of the results, it is important to keep in mind that the effective lagrangian on which these calculations are based, \textit{was fitted to ground-state properties of selected spherical nuclei in the stability region}~\cite{f.s.t.} and that there has been \textit{no refitting} of these parameters.

\begin{figure}
\begin{center}
\vspace{0.9cm}
\includegraphics[width=0.5\textwidth,height=0.3\textheight,clip=true]{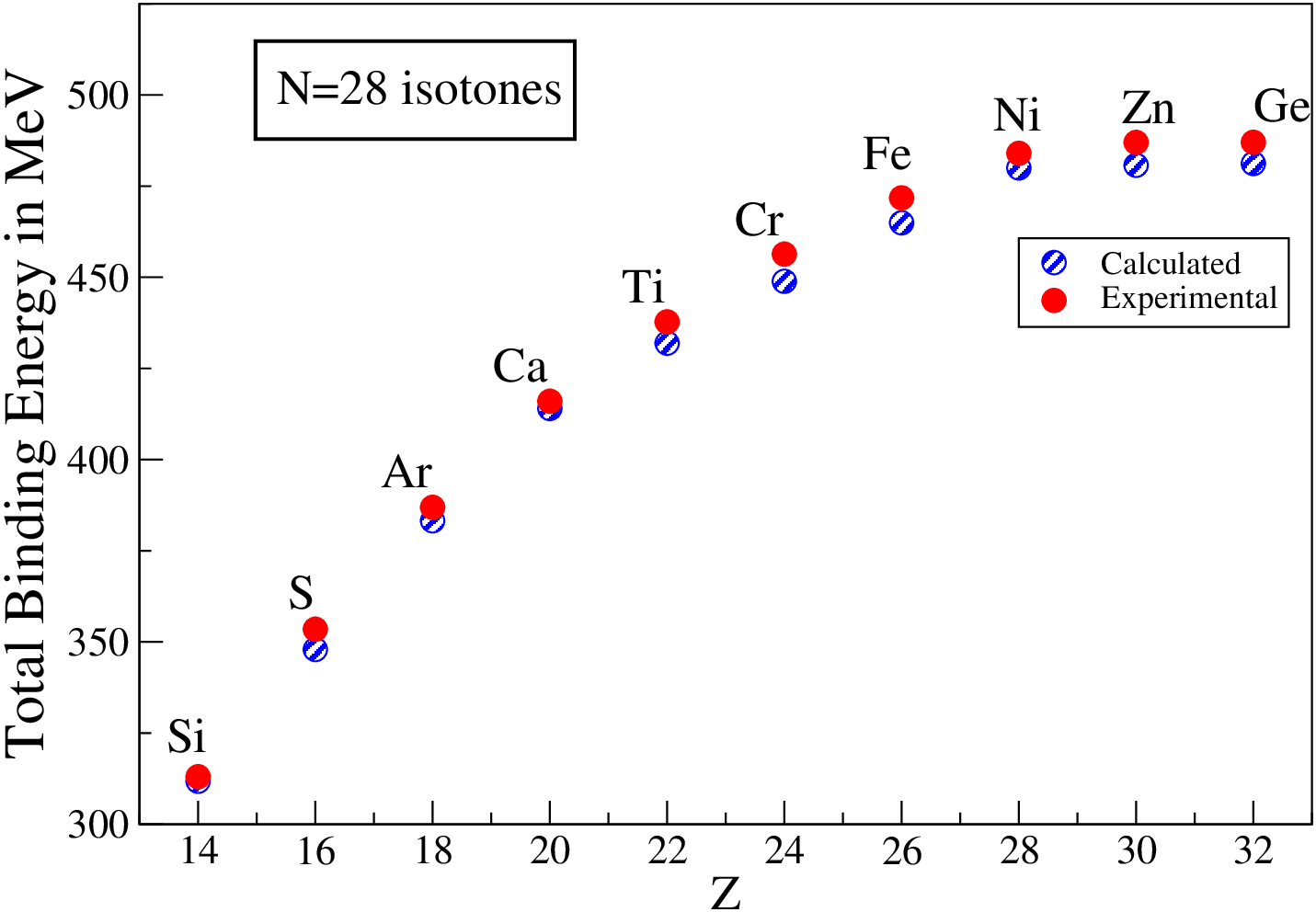}
\caption{\label{be-n28}Calculated and experimental values of the total binding energy for even N=28 isotones. Energies are positive and measured in MeV.}
\end{center}
\end{figure}

\begin{figure}
\begin{center}
\vspace{0.7cm}
\includegraphics[width=0.5\textwidth,height=0.3\textheight,clip=true]{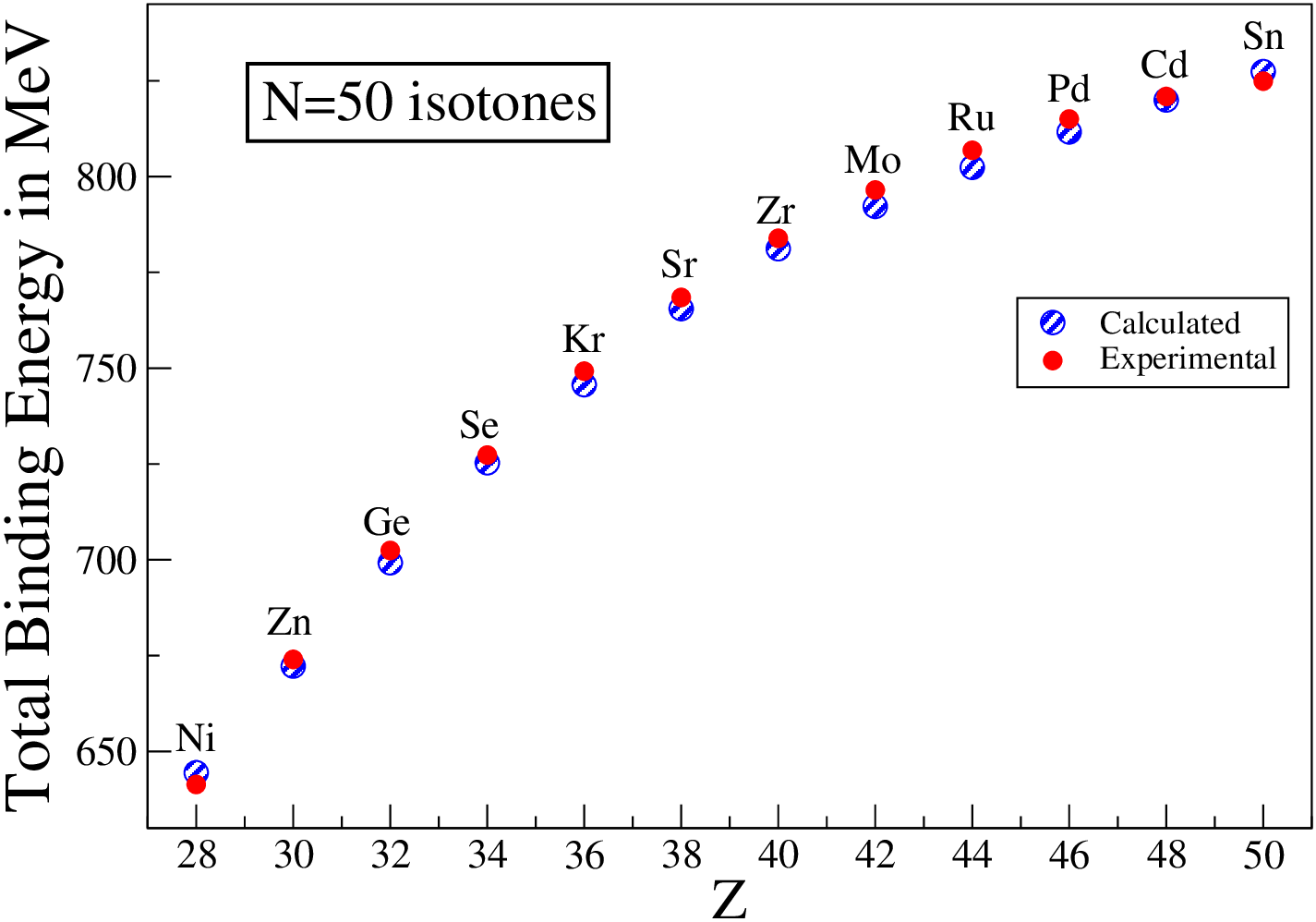}
\caption{\label{be-n50}Calculated and experimental values of the total binding energy for even N=50 isotones. Energies are positive and measured in MeV.}
\end{center}
\end{figure}

\begin{figure}
\begin{center}
\vspace{0.7cm}
\includegraphics[width=0.5\textwidth,height=0.3\textheight,clip=true]{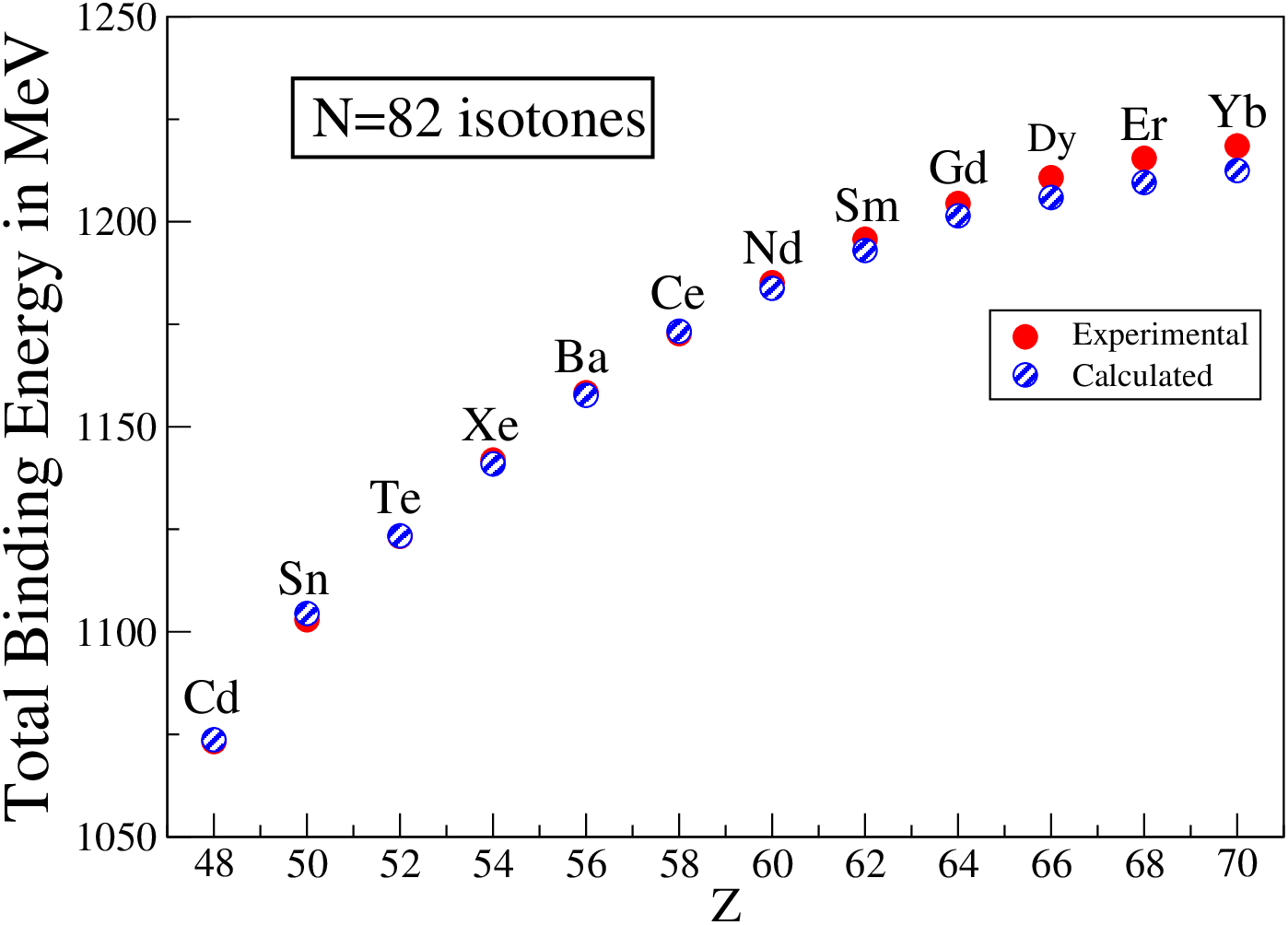}
\caption{\label{be-n82}Calculated and experimental values of the total binding energy for even N=82 isotones. Energies are positive and measured in MeV.}
\end{center}
\end{figure}

\begin{figure}
\begin{center}
\vspace{0.7cm}
\includegraphics[width=0.5\textwidth,height=0.3\textheight,clip=true]{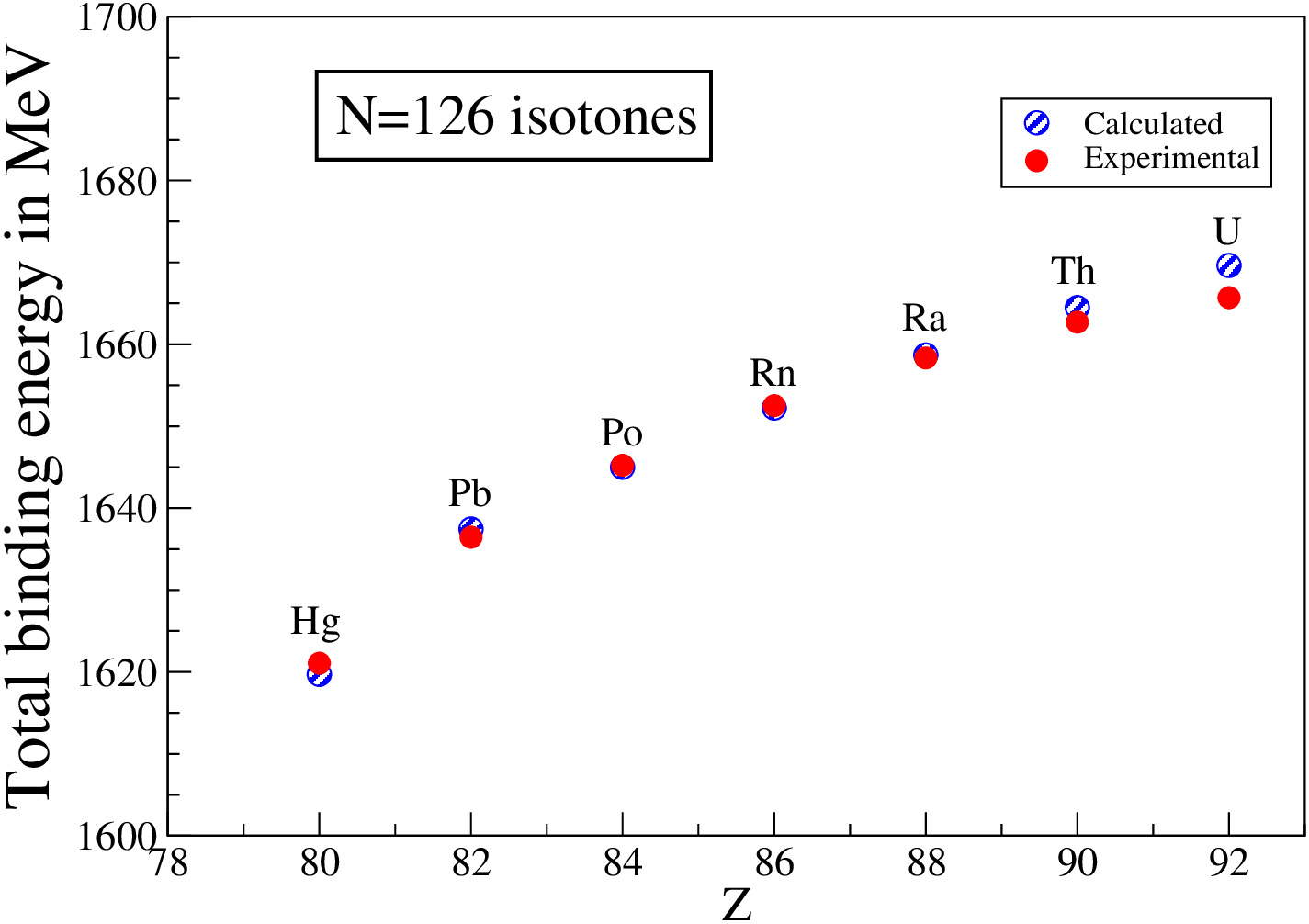}
\caption{\label{be-n126}Calculated and experimental values of the total binding energy for even N=126 isotones. Energies are positive and measured in MeV.}
\end{center}
\end{figure}

\begin{figure}
\begin{center}
\vspace{0.7cm}
\includegraphics[width=0.5\textwidth,height=0.3\textheight,clip=true]{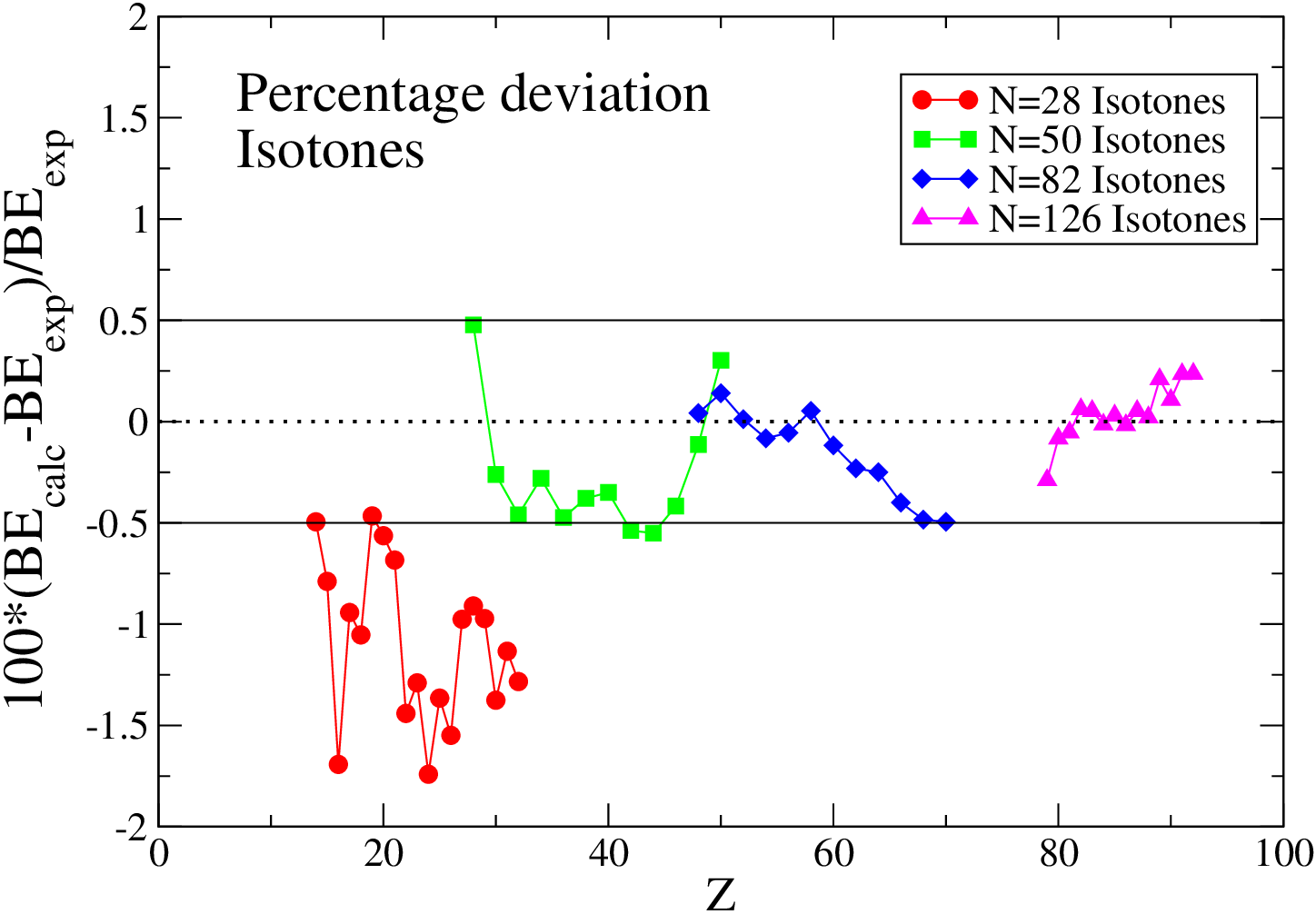}
\caption{\label{perc-isotones}Percentage difference between calculated and experimental values for all even isotones.}
\end{center}
\end{figure}

The first set of results shown correspond to the total binding energy of even-even isotones with total neutron number, N, equal to 28, 50, 82 and 126. These are shown in Fig.~\ref{be-n28} - Fig.~\ref{be-n126}. As can noted from these figures, there is a very good agreement between the calculated and experimental values, which are taken from~\cite{nudat}. In Fig.~\ref{perc-isotones}, the percentage deviation, or error, of the results is shown. The overall agreement can be seen to improve as one goes to larger nuclei and, except for the N=28 isotones, the percentage error is in general less than 0.5\%. 

%Place first set of plots here

\begin{figure}
\begin{center}
\vspace{1cm}
\includegraphics[width=0.5\textwidth,height=0.3\textheight,clip=true]{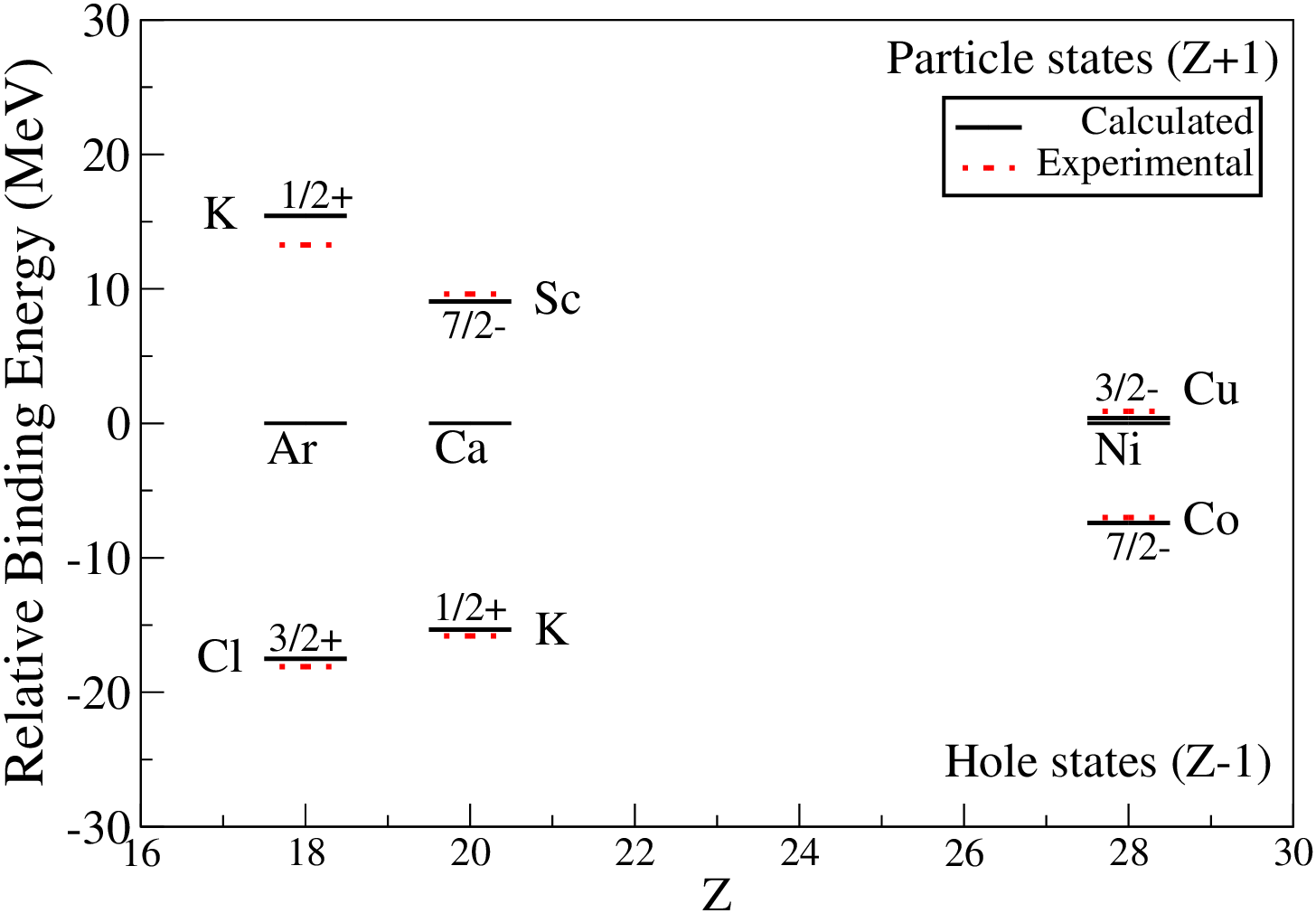}
\caption{\label{ph-n28}Particle and hole states neighboring even-even N=28 isotone nuclei. Binding energies are considered positive. Only nuclei with particle and hole states with respect to filled j-shells have been plotted. The values of Z correspond to the even-even nuclei in the middle row. Calculated J$^{\pi}$ values are included and, unless  indicated otherwise, are in agreement with those of the experimental ground state. Where the experimental J$^{\pi}$ value disagrees with the calculated one, or is uncertain, it is also included in the figure.}
\end{center}
\end{figure}

\begin{figure}
\begin{center}
\vspace{1cm}
\includegraphics[width=0.5\textwidth,height=0.3\textheight,clip=true]{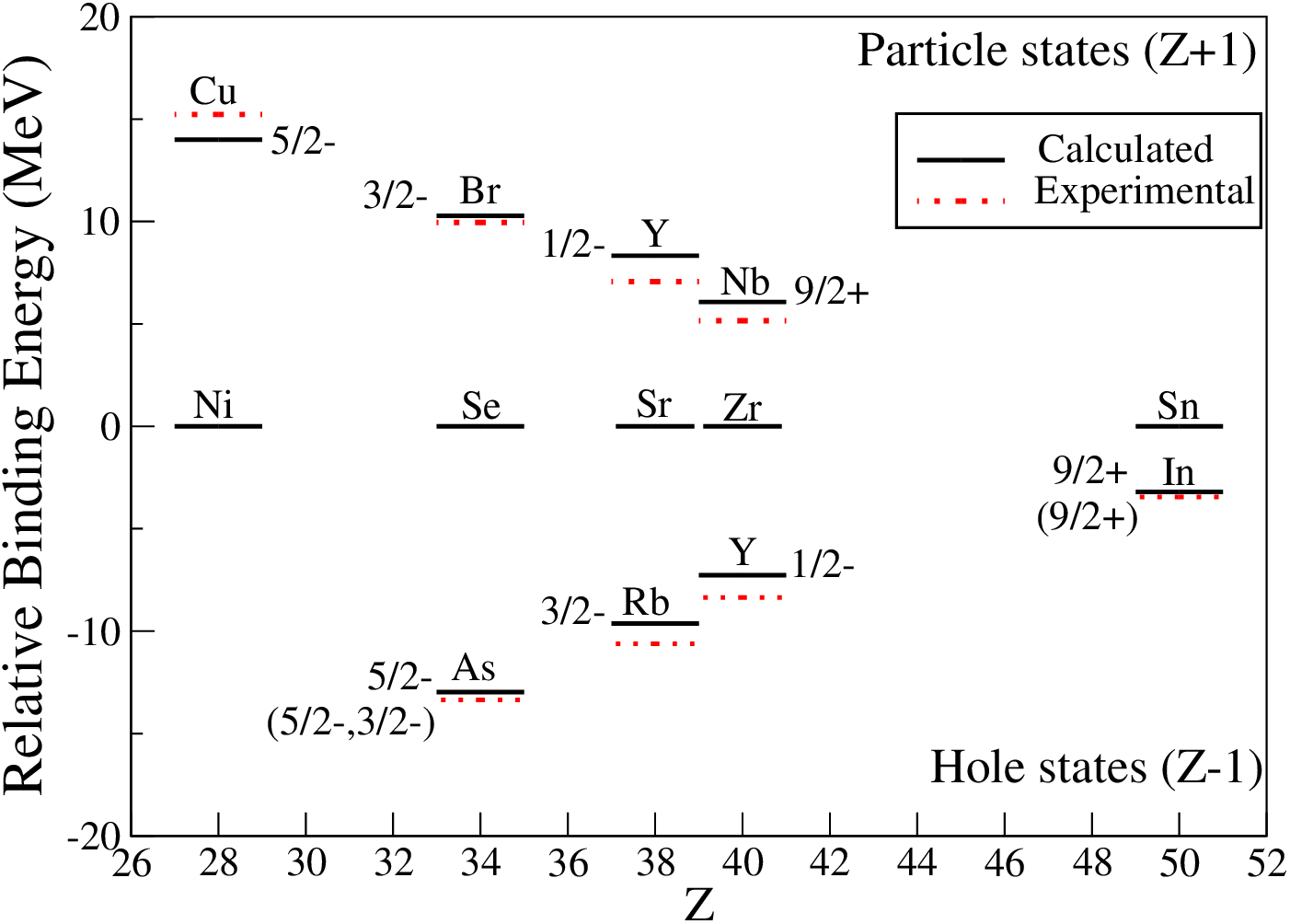}
\caption{\label{ph-n50}Same as in Fig.~\ref{ph-n28} for N=50 isotones.}
\end{center}
\end{figure}

\begin{figure}
\begin{center}
\vspace{1cm}
\includegraphics[width=0.5\textwidth,height=0.3\textheight,clip=true]{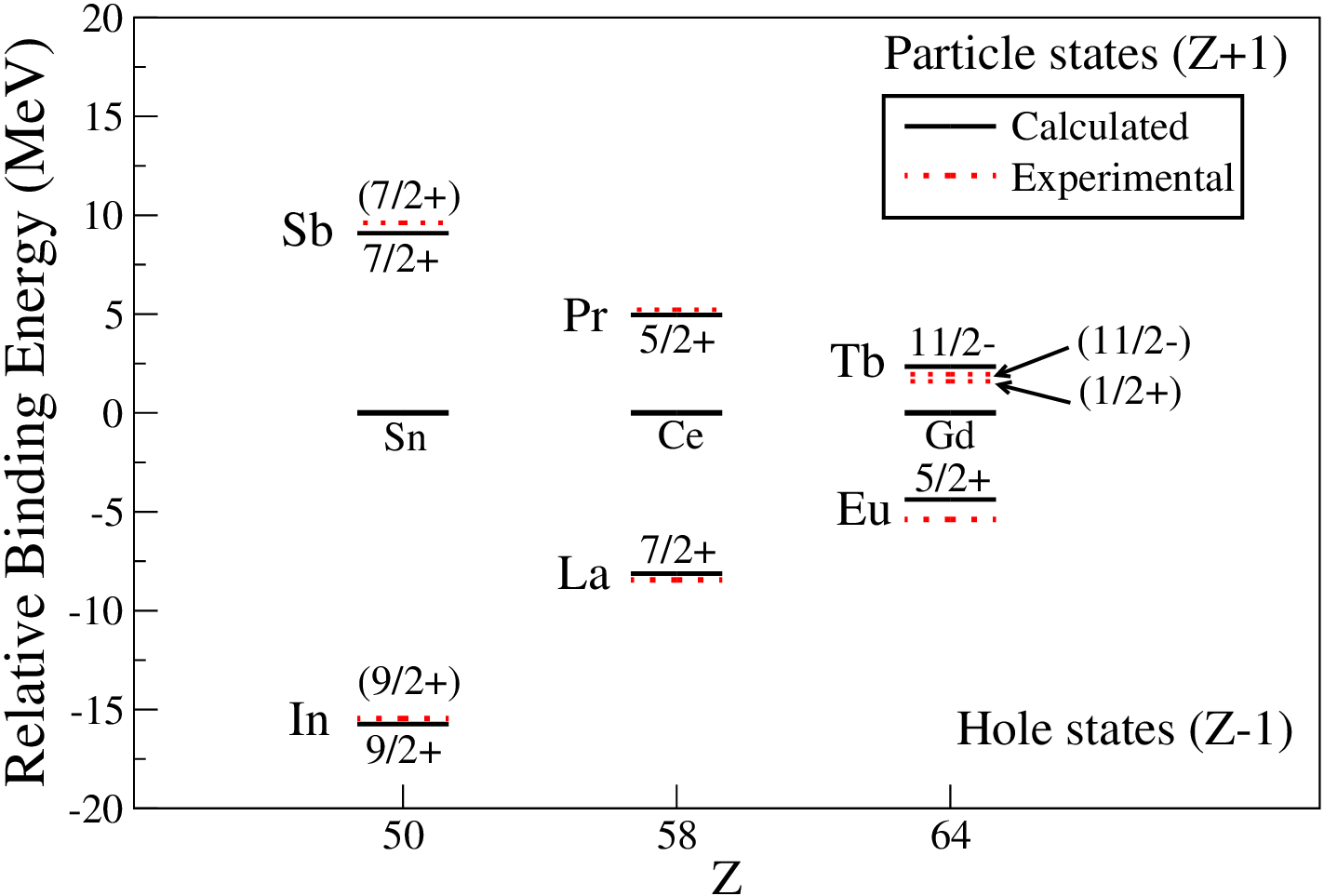}
\caption{\label{ph-n82}Same as in Fig.~\ref{ph-n28} for N=82 isotones. The position of the (11/2-) metastable state of $^{147}\textrm{Tb}$ has been shifted so it can be seen in this energy scale. It actually lies 50 keV above the ground-state.}
\end{center}
\end{figure}

\begin{figure}
\begin{center}
\vspace{1cm}
\includegraphics[width=0.5\textwidth,height=0.3\textheight,clip=true]{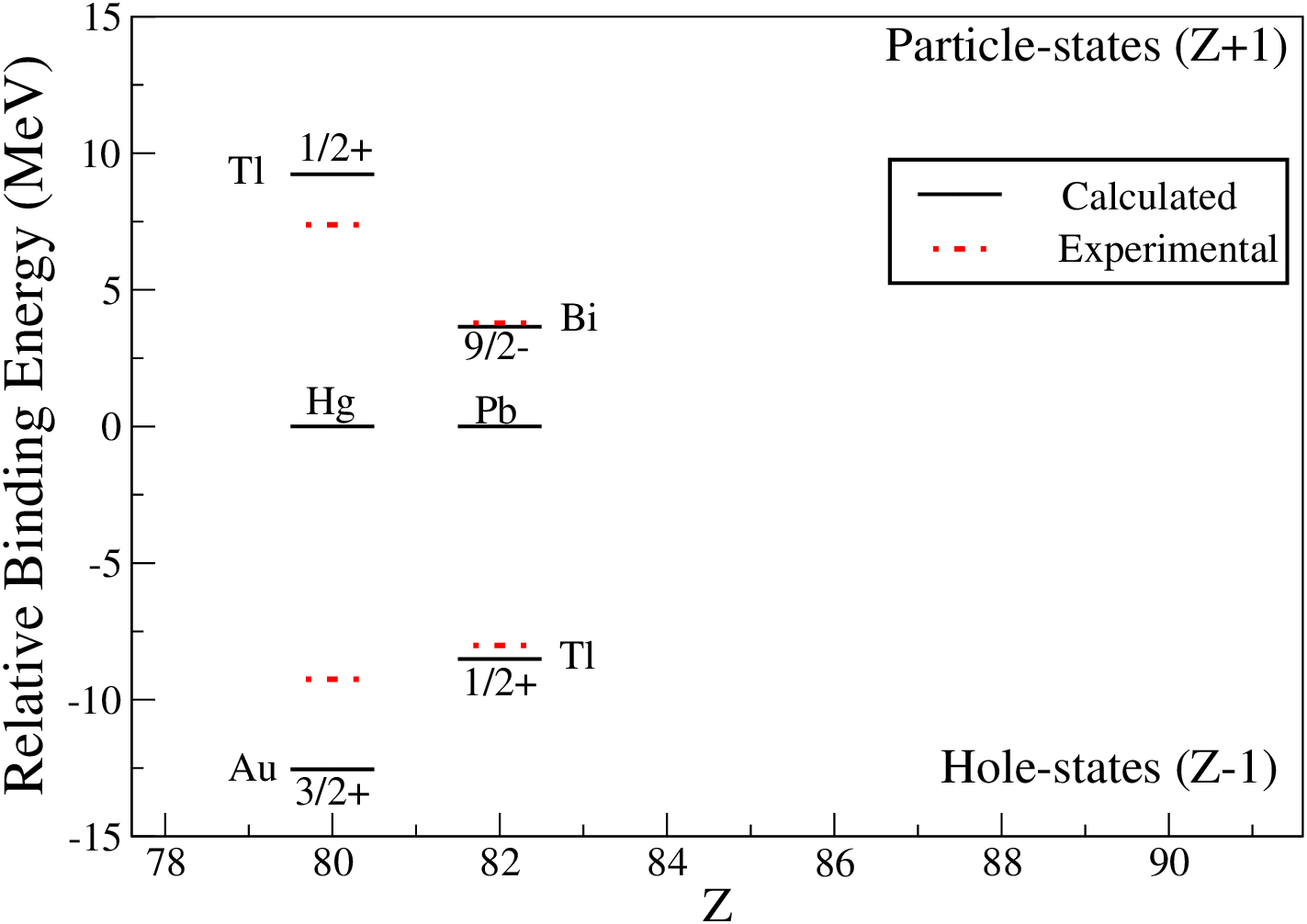}
\caption{\label{ph-n126}Same as in Fig.~\ref{ph-n28} for N=126 isotones.}
\end{center}
\end{figure}

The following set of figures, Fig.~\ref{ph-n28} - Fig.~\ref{ph-n126}, show the predictions for the ground-state energies of nuclei neighboring even-even isotones of given N. The energy differences are taken between the odd nucleus and the closest even nucleus. These differences in energy, i.e. the magnitude of the chemical potential, are calculated using the binding energies obtained in this EFT/DFT  approach and are compared with available experimental data. The even-even isotones  shown in these plots, and which are located at the zero point of energy, correspond to nuclei with (calculated) closed j-shells. A filled j-shell has an even number of particles, and J$^{\pi}$ equal to 0+, here only particles or holes relative to filled j-shells contribute to the total angular momentum and parity of the nucleus. This is precisely the argument given in the shell model~\cite{waleckabook}. Since these calculations are effectively mean-field calculations, there is no explicit inclusion of pairing~\cite{marco1}. Thus states corresponding to partially filled j-shells are not included in these plots.

%PLACE SECOND SET OF PLOTS HERE

From these plots one can appreciate the excellent agreement between the calculated and experimental values of this energy difference, or chemical potential. In addition to this, the predicted ground-state quantum numbers of the odd nuclei agree with the experimental values. The only disagreement in all the isotone cases considered occurs for $^{147}_{\ 65}\textrm{Tb}_{82}$, Fig.~\ref{ph-n82}. Here the experimental ground-state corresponds to a (1/2+) state while the predicted ground-state is 11/2-. The result is not far from the metastable state (11/2-) which lies 50 keV above the true ground-state~\cite{levels-Tb147}. This is indicated in Fig.~\ref{ph-n82}, where the metastable state has been shifted in order to be seen on this energy scale. 

\begin{figure}
\begin{center}
\vspace{1cm}
\includegraphics[width=0.5\textwidth,height=0.3\textheight,clip=true]{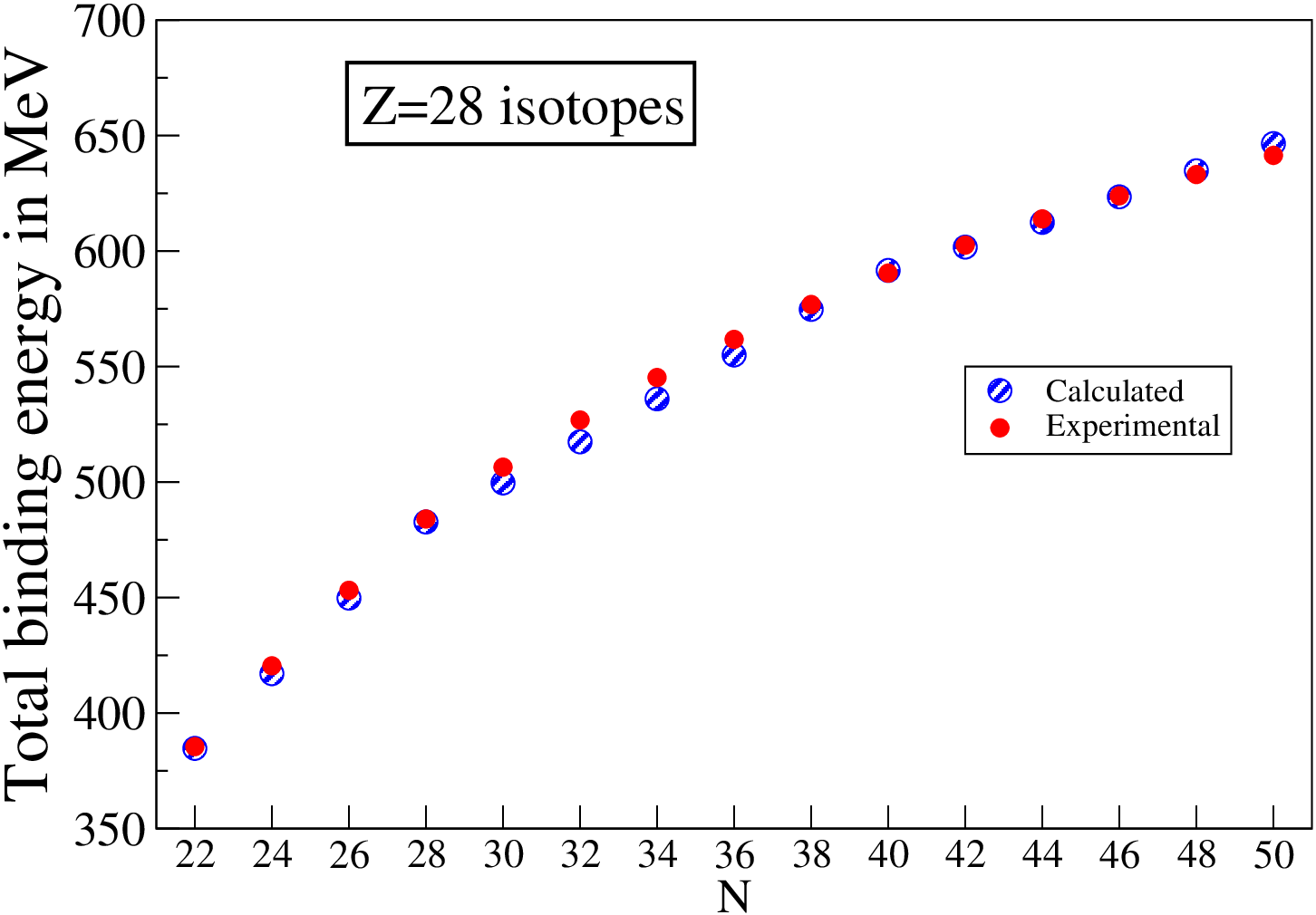}
\caption{\label{be-z28}Calculated and experimental values of the total binding energy for even Z=28 isotopes. Energies are positive and measured in MeV.}
\end{center}
\end{figure}

\begin{figure}
\begin{center}
\vspace{1cm}
\includegraphics[width=0.5\textwidth,height=0.3\textheight,clip=true]{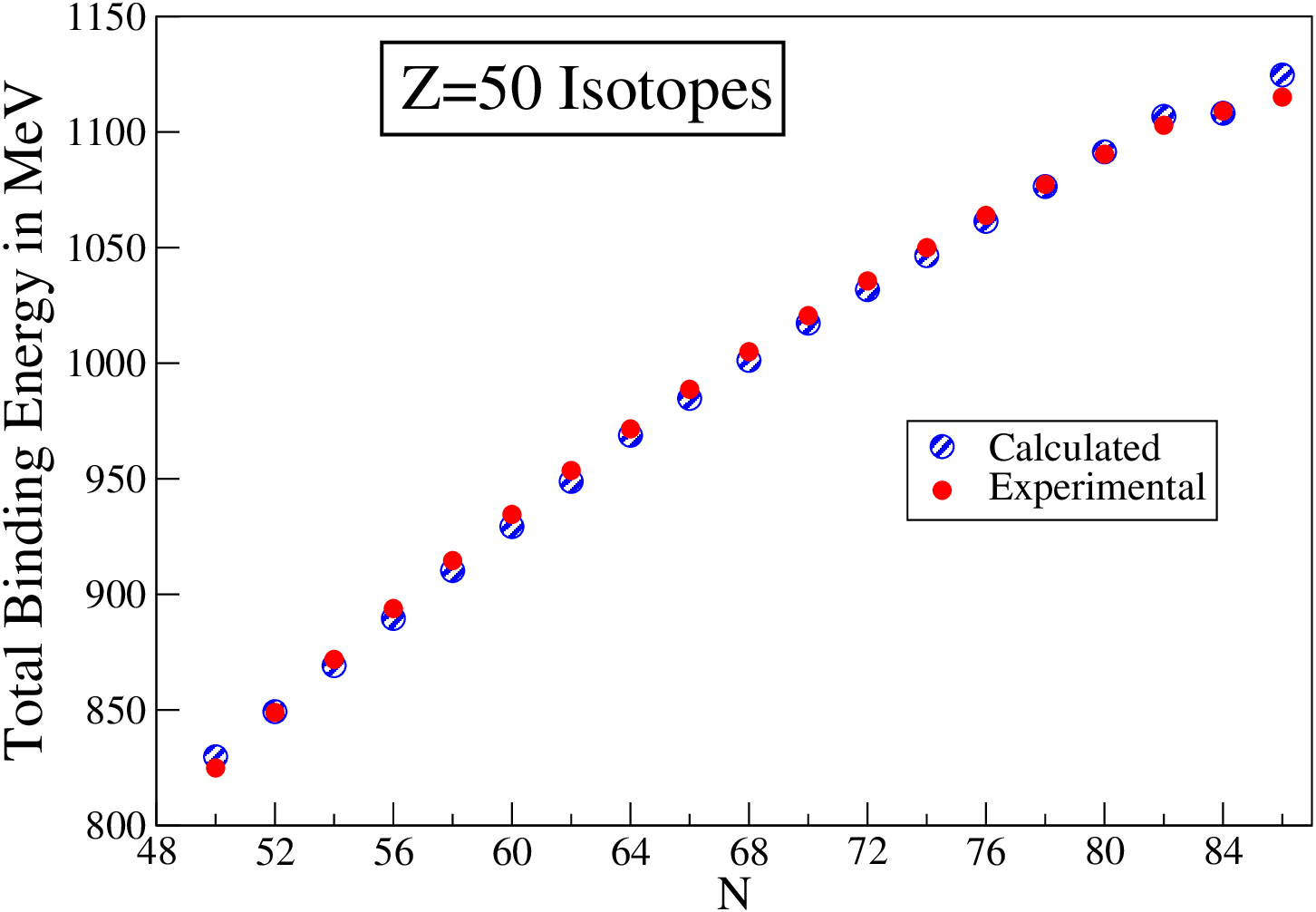}
\caption{\label{be-z50}Calculated and experimental values of the total binding energy for even Z=50 isotopes. Energies are positive and measured in MeV.}
\end{center}
\end{figure}

\begin{figure}
\begin{center}
\vspace{1cm}
\includegraphics[width=0.5\textwidth,height=0.3\textheight,clip=true]{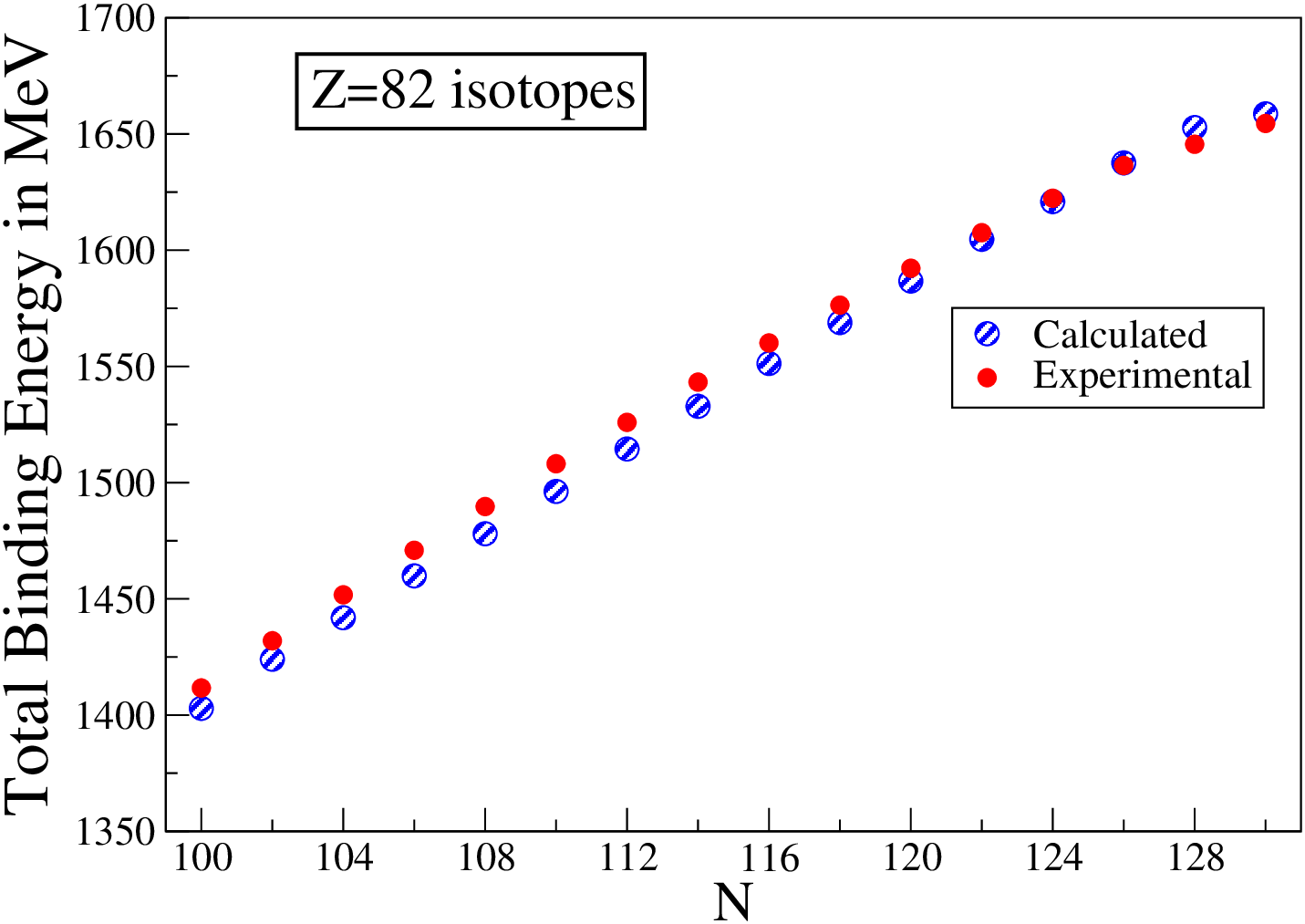}
\caption{\label{be-z82}Calculated and experimental values of the total binding energy for even Z=82 isotopes. Energies are positive and measured in MeV.}
\end{center}
\end{figure}

\begin{figure}
\begin{center}
\vspace{1cm}
\includegraphics[width=0.5\textwidth,height=0.3\textheight,clip=true]{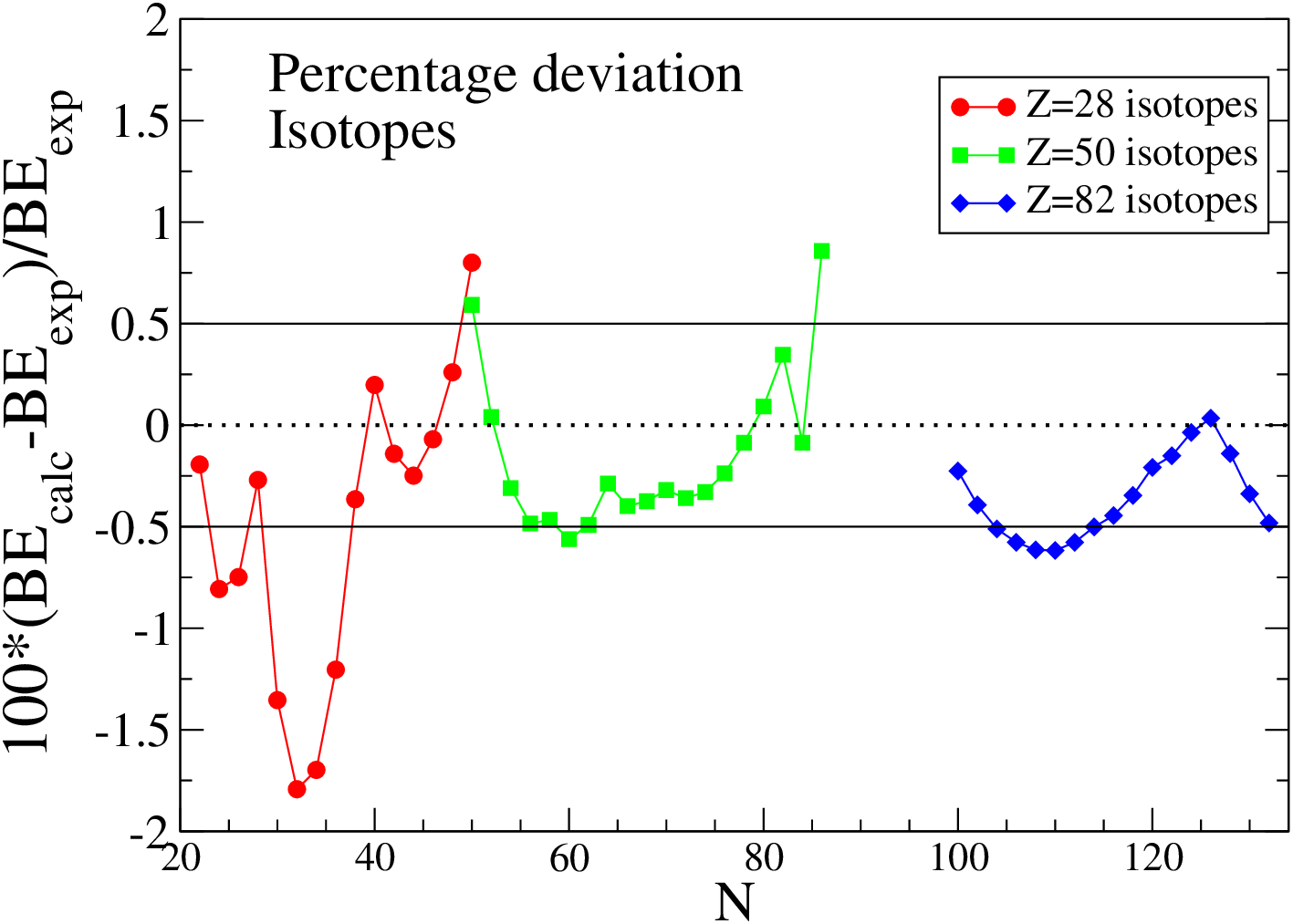}
\caption{\label{percentage-isotopes}Percentage difference between calculated and experimental values for all even isotopes considered.}
\end{center}
\end{figure}

The next set of plots, Fig.~\ref{be-z28} - Fig.~\ref{be-z82}, show the ground-state total binding energy of even-even isotopes with Z= 28, 50 and 82. They are equivalent to the plots shown for the isotones. Here again the predicted values agree overall with experiment below the 1\% level as can be seen in Fig.~\ref{percentage-isotopes}. The agreement between predicted and experimental data improves as one moves to larger nuclei.

%PLACE THIRD SET OF PLOTS HERE

\begin{figure}
\begin{center}
\vspace{1cm}
\includegraphics[width=0.5\textwidth,height=0.3\textheight,clip=true]{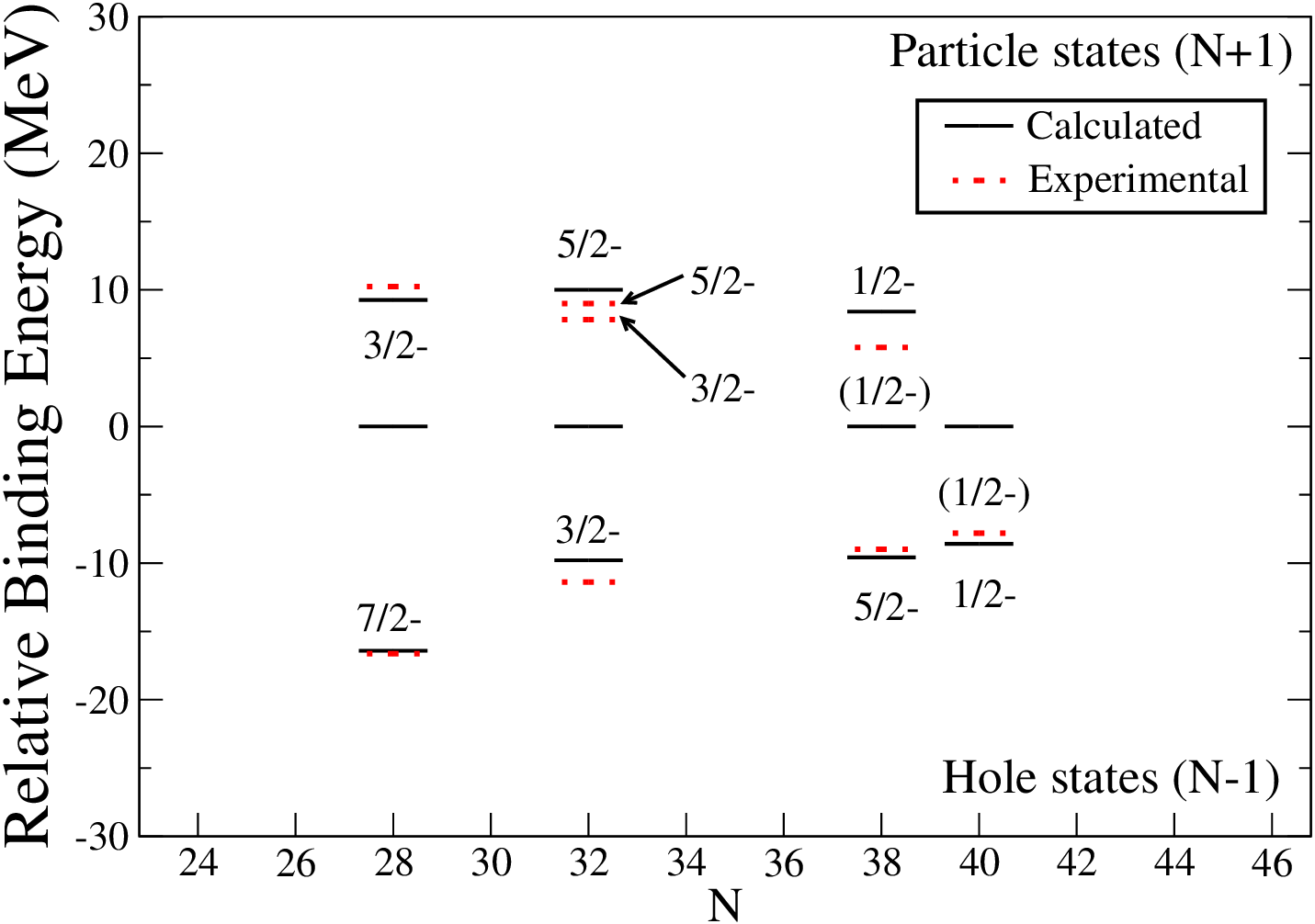}
\caption{\label{ph-z28}Particle and hole states neighboring even-even Z=28 isotopes. Binding energies are considered positive. Only nuclei with particle and hole states with respect to calculated filled j-shells have been plotted. The value of N corresponds to the even-even isotopes in the middle of row. Calculated J$^{\pi}$ values are included and, unless indicated otherwise, are in agreement with the experimental ground states. Where there is disagreement with the experimental J$^{\pi}$ value, or it is uncertain, the experimental ground-state J$^{\pi}$ value is also indicated in the plot. In addition, in this case, the lowest-lying excited state with the calculated J$^{\pi}$ is also included.}
\end{center}
\end{figure}

\begin{figure}
\begin{center}
\vspace{1cm}
\includegraphics[width=0.5\textwidth,height=0.3\textheight]{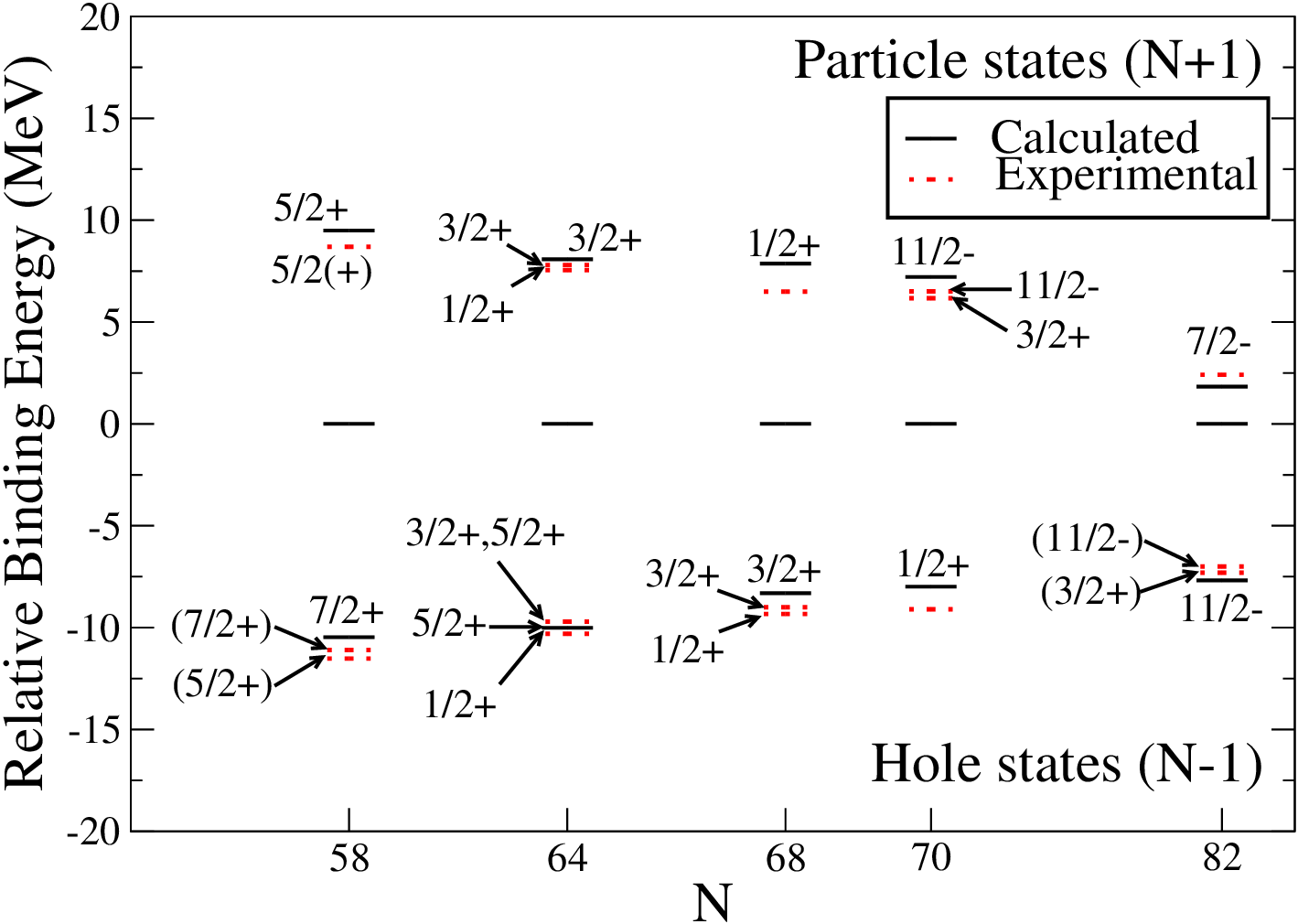}
\caption{\label{ph-z50}Same as in Fig.~\ref{ph-z28} for Z=50 isotopes.}
\end{center}
\end{figure}

\begin{figure}
\begin{center}
\vspace{1cm}
\includegraphics[width=0.5\textwidth,height=0.3\textheight]{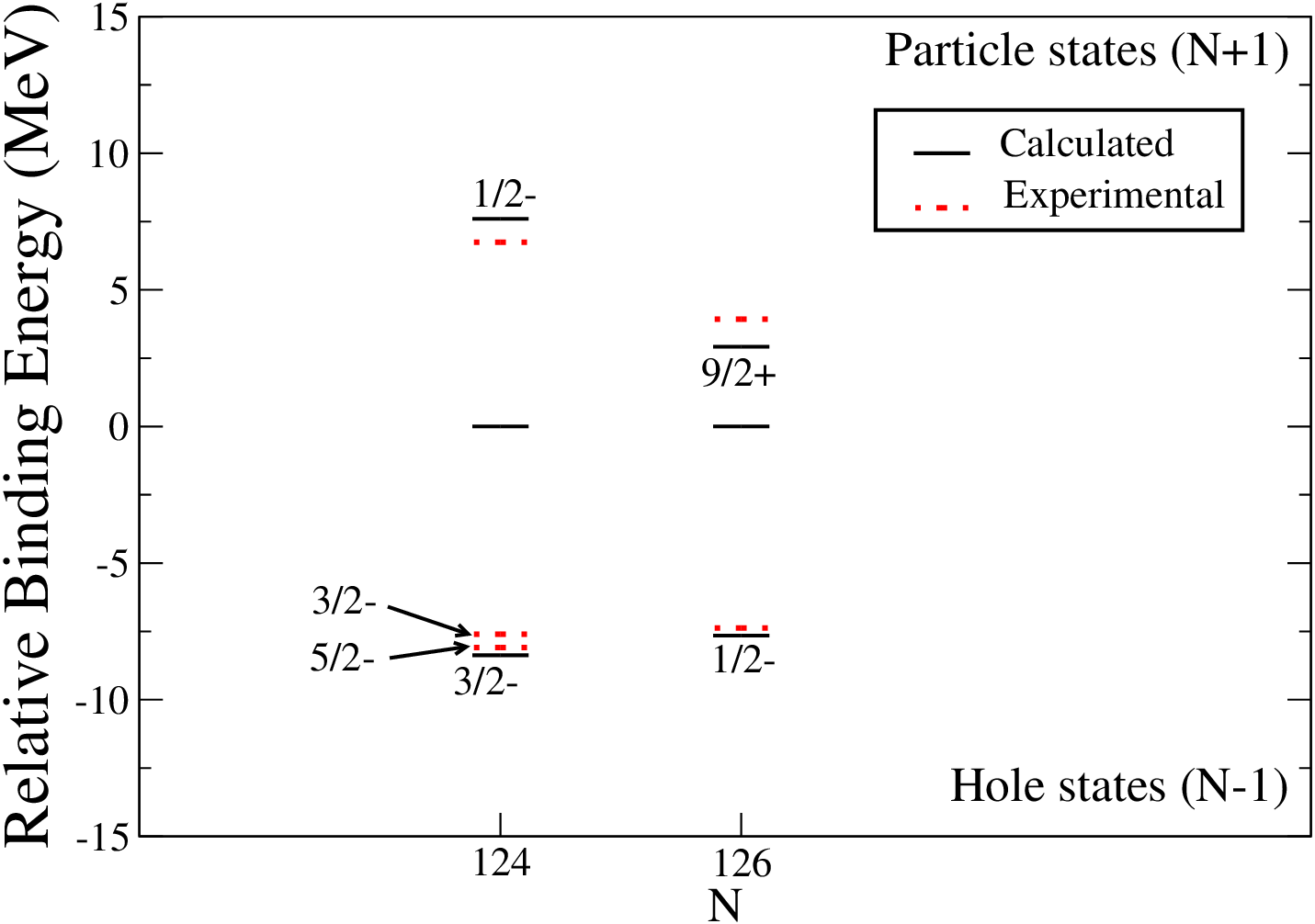}
\caption{\label{ph-z82}Same as in Fig.~\ref{ph-z28} for Z=82 isotopes.}
\end{center}
\end{figure}

The final set of plots, Fig.~\ref{ph-z28} - Fig.~\ref{ph-z82}, correspond to predictions of the relative binding energy of odd nuclei neighboring even-even isotopes of given Z, with respect to the binding energy of these even-even nuclei. As in the equivalent plots for isotones, the even-even nuclei correspond to the calculated filled j-shells and locate the zero-energy level. 

%PLACE FOURTH SET OF PLOTS HERE

Figure ~\ref{ph-z28} shows the results for isotopes of Ni (Z=28). Here one can observe that overall, there is a very good agreement with experiment, similar to that of the N=28 isotones. In the case of  $^{61}_{28}\textrm{Ni}_{33}$, there is a disagreement in the predicted quantum numbers of the ground-state. The experimental value corresponds to a 3/2- state, while the predicted one is a 5/2-. There is an excited state with J$^{\pi}$ value in agreement with the calculated ground-state that lies only 70 keV above it~\cite{levels-ni61}. In this figure, this separation in energy has been exaggerated to show it in the scale of the plot. The disagreement in the location of the ground-state shows the sensitivity of the actual ordering of states to the form of the central potential. One can understand this by looking at the effects of putting additional neutrons into the system. There is a major shell closure at N=28. If one more neutron is added to the system it will go into the next j-shell which is $j=3/2$ corresponding to a 2p3/2 level, as expected from the naive shell model~\footnote{The level ordering is sensitive to the actual form of the central potential, and here we use the one presented in~\cite{waleckabook}, chapter 6.}. This level becomes completely filled at N=32, since $2j+1 = 4$, which is correctly predicted as can be seen in the plot. Adding one more particle to this nucleus, should start filling the next j-shell, the 1f5/2 level, giving the ground state the quantum numbers 5/2- as predicted. Instead, it seems that it is energetically more favorable to promote one particle from the 2p3/2 level to the 1f5/2 leaving a hole. The two particles in the 1f5/2 level do not contribute to the angular momentum, which is solely determined by this hole in the 2p3/2. Adding more neutrons fills the 1f5/2 until it is completely full at N=38, since $2j+1 = 6$, which again is correctly predicted. After this, more neutrons just start filling the next level, 2p1/2, following the ordering given by the naive shell model. Therefore the predictions follow very closely the way the neutron levels are being filled and the fact that the disagreement at N=33 is small only shows how sensitive the ordering is to the form of the central potential.

The exact ordering of energy levels has a larger effect when the level density is high and the levels are very close together. This is precisely what happens in the case of isotopes with large values of Z. For any given Z the density level of neutrons close to the Fermi surface is larger since in general N$\geq$Z making the energy levels very close to each other. This can be appreciated in Fig.~\ref{ph-z50} where, if there is a disagreement between the calculated and experimental ground-state value of J$^{\pi}$, the lowest-lying excited state with the calculated quantum numbers has also been plotted~\cite{levels-sn}. Note in particular the small ($\approx$ tens of keV) energy differences between these states and each case and the fact that there is always a partner with quantum numbers that agree with the ones predicted for the ground-state. As for similar cases before, the energy difference has been exaggerated to be appreciated at this scale. 

A similar effect occurs in the case of the isotopes of Pb (Z=82). Figure~\ref{ph-z82} shows the filled j-shell states closest to $^{208}_{\ 82}\textrm{Pb}_{126}$ which is a mayor shell closure. Adding an extra neutron to this nucleus will start filling the next j-shell, which according to the naive shell model is a 2g9/2 state. This is precisely what is predicted here as can be seen in this figure for the particle state at N=126 (i.e. N=127). Removing a neutron from $^{208}_{\ 82}\textrm{Pb}_{126}$ corresponds to depleting the 3p1/2 state, as can be seen in the figure, and again it is also correctly predicted. Since $^{208}_{\ 82}\textrm{Pb}_{126}$ formed part of the nuclei used to fit the parameters of the theory, it is expected that there will be an accurate description of the neighboring nuclei, and in fact there is. 

Removing one additional neutron completely depletes the $j=1/2$ shell bringing the total number of neutrons to N=124, which is also observed occurring in this plot. According to~\cite{waleckabook} (see footnote above) the next hole level would be a 1i13/2 level, i.e. 13/2+, but the experimental evidence indicates that this level is in fact a 5/2-. The reordering of the levels, with respect to the naive shell model, can be understood as either a large splitting between the 3p1/2 and 3p3/2 levels, which brings the 3p1/2 level above the 1i13/2, or a considerable lowering of the 1i13/2 which brings the 1i13/2 below the 3p1/2. In either case removing one more neutron from this nucleus will start depleting the next j-shell which could be a 3p3/2 or a 2f5/2. In Fig.~\ref{ph-z82} it is shown that this level is a 5/2- in disagreement with the calculated one,  3/2-. The closest excited state having this predicted quantum numbers lies only 262 keV above the ground-state. This state is also plotted~\cite{levels-pb123}. The energy difference has been exaggerated in the figure, as in the previous plots. The reordering between the 2f5/2 and 3p3/2, shift the position of the real j-shell closures with respect to the calculated ones, as happened for the Z=50 isotopes shown in Fig.~\ref{ph-z50}; therefore they have not been included in this plot. Despite this disagreement, the magnitude of the chemical potentials of these nuclei are relatively well described.

In summary, these results show that the binding energy of all even-even nuclei with semi-magic numbers of neutrons and protons, N = 28, 50, 82, 126 and Z = 28, 50, 82 are reproduced to 1\% or better in this EFT/DFT approach with parameters G1 fit along the valley of stability. With all the isotones, the chemical potential of nuclei with a single proton particle or hole outside of filled j-shells is well reproduced and the predicted value of J$^{\pi}$ for the ground-state of these nuclei agrees with experiment.

When the number of neutrons gets very large in the regime where N $>$ Z, the level density is very high and the detailed ordering of the levels between major shells depends sensitively on the details of the various densities. Thus, while the chemical potential (binding energies) are still relatively well-described, the theory begins to lose its predictive power for the spin and parity of the ground states of single neutron particle and hole nuclei neighboring those with filled j-shells.

\section{Conclusions.}

This work investigates the applicability of the EFT/DFT approach to reproduce \textit{total binding energies} and \textit{chemical potentials} of semi-magic nuclei (N = 28, 50, 82, 126 and Z= 28, 50, 82) far from the valley of stability.  It extends the work done in~\cite{marco1}, where the convergence of this EFT/DFT approach was studied for doubly-magic nuclei. All calculations were performed using an independently developed program to solve the Kohn-Sham equations derived from the most sophisticated effective lagrangian given in~\cite{f.s.t.} with parameter set G1.

From the figures given in the previous section, several conclusions are drawn:

\begin{enumerate}
\item The EFT/DFT approach can reproduce the total binding energies of even-even isotones and isotopes with semi-magic numbers of neutrons and protons, respectively. The agreement is in general better than 1 \% where the best results are obtained for larger nuclei. This conclusion is in agreement with the previous results obtained in~\cite{marco1}.
\item The predictions for the magnitude of the chemical potentials for odd nuclei corresponding to particle or holes neighboring filled j-shells agree overall with the experimental evidence. This agreement varies for different nuclei and on average lies between 10\% and 20\%.
\item Regarding the predictions of the ground-state quantum numbers of odd nuclei neighboring even-even nuclei with filled j-shells, the results indicate that in the case of isotones these are accurately reproduced; for heavy isotopes with N $>$ Z, although the predictions of chemical potentials are in essential agreement with experiment, the EFT/DFT approach begins to lose its predictive power.
\end{enumerate}

Overall, the use of this effective field theory approach for QCD and the interpretation of the resulting relativistic Hartree equations through Kohn-Sham potentials and density functional theory~\cite{f.s.t.} provides an excellent description of the ground-state properties of semi-magic nuclei when extrapolated away from the valley of stability.

\section*{acknowledgments}
%put your acknowledgments here.
The author will like to thank Dr. J. D. Walecka for his support and advice. This work is supported in part by DOE grant DE-FG02-97ER41023.

\end{document}